\documentclass[preprint,superscriptaddress,showpacs,preprintnumbers,amsmath,amssymb]{revtex4-1}

\usepackage{amsmath}
\usepackage{amssymb}
\usepackage{graphicx}
\usepackage{morefloats}
\usepackage[usenames,dvipsnames]{xcolor}
\usepackage{blkarray}
\usepackage{verbatim}
\usepackage{hyperref}
\usepackage{color}

\begin{document}

\title{Conformity-Driven Agents Support Ordered Phases in the Spatial Public Goods Game}

\author{Marco Alberto Javarone}
\email{marcojavarone@gmail.com}
\affiliation{Department of Mathematics and Computer Science, University of Cagliari, Cagliari (Italy)}% Via Universita 40, 09124 Cagliari (Italy)}
\author{Alberto Antonioni}
\affiliation{University of Lausanne, Lausanne (Switzerland)}
\affiliation{Universidad Carlos III de Madrid, Legan\'es (Spain)}% Calle Madrid, 133, 28903 Getafe, Madrid (Spain)}
\affiliation{Institute for Biocomputation and Physics of Complex Systems, Zaragoza (Spain)}
%Mariano Esquillor, Edificio I + D - 50018 Zaragoza (Spain)}
\author{Francesco Caravelli}
\affiliation{Invenia Labs, Cambridge (UK)}%, 27 Parkside Place, Parkside, Cambridge CB1 1HQ, UK}
\affiliation{London Institute of Mathematical Sciences, London (UK)}%, 35a South Street, London W1K 2XF, UK}
%\affiliation{Department of Computer Science, University College London, Gower Street, London WC1E 6BT, UK}

\date{\today}

\begin{abstract}
We investigate the spatial Public Goods Game in the presence of fitness-driven and conformity-driven agents.
This framework usually considers only the former type of agents, i.e., agents that tend to imitate the strategy of their fittest neighbors.
However, whenever we study social systems, the evolution of a population might be affected also by social behaviors as conformism, stubbornness, altruism, and selfishness.
Although the term evolution can assume different meanings depending on the considered domain, here it corresponds to the set of processes that lead a system towards an equilibrium or a steady-state.
We map fitness to the agents' payoff so that richer agents are those most imitated by fitness-driven agents, while conformity-driven agents tend to imitate the strategy assumed by the majority of their neighbors.
Numerical simulations aim to identify the nature of the transition, on varying the amount of the relative density of conformity-driven agents in the population, and to study the nature of related equilibria. Remarkably, we find that conformism generally fosters ordered cooperative phases and may also lead to bistable behaviors.
\end{abstract}

\maketitle

In the last two decades, Evolutionary Game Theory~\cite{perc01,nowak01,nowak02,tomassini01,moreno01,solnperc01} (EGT) has strongly developed into a mature field that, nowadays, represents a vivid and independent research area with a list of applications spanning from biology to social systems~\cite{nowak02,nowak03,perc01,perc02,friedman01,shuster01,frey01}. 
In particular, EGT constitutes a powerful tool for modelling a myriad of natural phenomena characterized by evolutionary dynamics~\cite{hofbauer01} and complex interaction patterns~\cite{sayama15}.
The emergence of cooperation~\cite{axelrod84} is one of the central topics in EGT. The Prisoner's Dilemma~\cite{nowak02} (PD)
and its $n$-person PD, also called Public Goods Game,
represent the typical game-theoretical framework in which individual and group interests collide. In fact, in the PD
game two players can decide to mutually cooperate for achieving a common goal and, at the same time, are both tempted to exploit their opponent in order to obtain an higher payoff at the expense of the other player.
Among other mechanisms~\cite{nowak04}, \emph{network reciprocity} has been proposed as a fundamental tool to promote cooperation in structured populations of unlabeled and anonymous individuals.
Recently, a series of works~\cite{meloni01,tomassini02,perc02} unveiled that, even when agents' interactions are based upon Nash Equilibria~\cite{game01} in which \textit{defection} (not cooperation) theoretically dominates cooperative strategies, agent, and even human, populations are able to attain a cooperative equilibrium~\cite{meloni01,perc03,tomassini02,tomassini03,tomassini04,javarone01,javarone02,antonioni14}. 
In this context, it is interesting to identify behaviors and properties that may lead a population towards cooperation, although defection is risk-less and constitutes, egoistically, the most convenient strategy. For instance, random or purposeful motion~\cite{aktipis,vainstein,meloni01,tomassini02,javarone02} and competitiveness~\cite{javarone01} are able to increase the level of cooperation in the spatial Prisoner's Dilemma.
Remarkably, EGT allows also to study the role of peer influences in social dilemmas, as reported in~\cite{perc04,pena01}. In~\cite{perc04} the authors investigate the role of conformism~\cite{aronson01} in social dilemmas, by simulating a two-species population whose agents, embedded in a scale-free network~\cite{caldarelli01}, play pairwise the PD game with their neighbors; the two species forming the population were payoff-driven agents and conformist agents. Their result shows that conformism can enhance social reciprocity highlighting its beneficial role in the resolution of social dilemmas. In~\cite{pena01}, the authors report that scale-free networks reduce the amplification of cooperation when an opportune amount of conformists is introduced in the imitation rules.
It is worth observing that conformism is one of the most investigated behaviors in the field of sociophysics~\cite{galam01,loreto01}, i.e, the field attempting to analyze socio-economic systems under the lens of statistical physics~\cite{galam01,galam02}. 
Moreover, conformism has been shown to deeply influence opinion dynamics~\cite{galam02,javarone04,javarone05}, and more in general phenomena of social spreading, as well as evolutionary dynamics~\cite{perc04,javarone06}.
For instance, in~\cite{javarone05,javarone06} the authors show that conformism (and nonconformism) can lead a population towards a disordered phase, i.e., a coexistence of different agents, and towards non-linear dynamics (e.g. bistability).

In the present work, we investigate the spatial Public Goods Game (PGG hereinafter) by considering a population composed of conformity-driven agents and fitness-driven agents. Thus, while the former tend to update their strategy with the most adopted one in their neighborhood, the latter tend to imitate their richest neighbor (i.e., the most fitted).
Although the PGG exhibits a theoretically predicted Nash Equilibrium of defection, previous works identified several strategies to support cooperation, spanning from awarding mechanisms (e.g.~\cite{perc03}) to optimal game settings. Notably, in~\cite{perc03} authors report that the synergy factor, usually indicated as $r$, adopted to compute the agents' payoff, can be opportunely tuned in order to support cooperation on bi-dimensional regular lattices. This result is very important as it entails that if the payoff of cooperators reaches, or overtakes, a minimum value, all agents turn their strategy to cooperation. As below, the minimum threshold of the synergy factor depends on the topology of the population (i.e., the way agents are arranged).
Therefore, adding a social influence in the PGG implies dealing with two degrees of freedom: the synergy factor $r$ (whose individual effect is known) and the density of conformist agents $\rho_c$.
The proposed model is studied by means of numerical simulations and performed by arranging agents that play a spatial PGG on a bi-dimensional regular lattice with periodic boundary conditions.

We now introduce the model studied in this Letter. In general, the PGG considers a population of $N$ agents that can adopt two different strategies: cooperation and defection. At each time step, cooperators provide a unitary contribution to a common pool, whereas defectors do the opposite, i.e., not contribute.
Then, the payoff of a cooperator (i.e., $\pi^{c}$) and that of a defector (i.e., $\pi^{d}$) read
\begin{equation}\label{eq:pgg_payoff}
\begin{cases}
\pi^{c} = r \frac{N^c}{G} - c\\
\pi^{d} = r \frac{N^c}{G}
\end{cases}
\end{equation}
\noindent where $N^c$ is the number of cooperators among the $G$ agents involved in the game (i.e., the considered agent with its neighbors), $r$ synergy factor, and $c$ agents' contribution (we set to $1$ for all cooperators, without the loss of generality). As we discuss below, the value of $G$ depends on the topology adopted to arrange the agent population.
\begin{figure*}[t]
\centering
\includegraphics[width=0.8\textwidth]{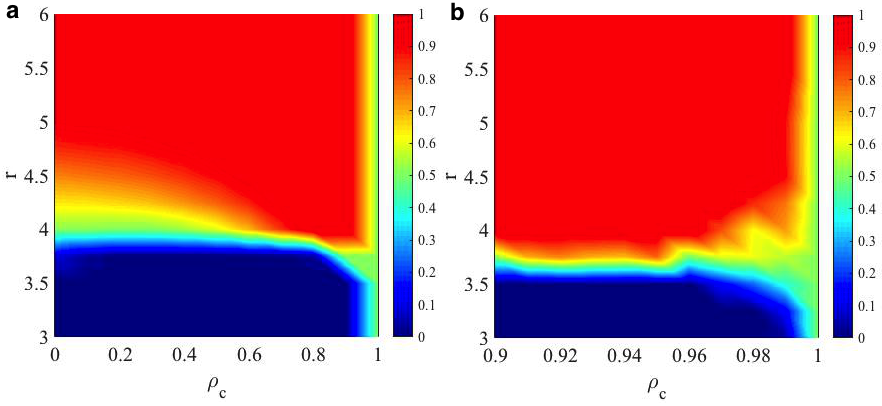}
\caption{\small (Color online) Cooperation diagram on varying $\rho_c$ in a population with $N = 10^{4}$. \textbf{a} $\rho_c$ in range $\in [0.0,1.0]$. \textbf{b} $\rho_c$ in range $\in [0.9,1.0]$. Red corresponds to areas of cooperation, while blue to those of defection. Results are averaged over $50$ simulation runs and have been 
computed using $11\times 11$ parameter values. \label{fig:cooperation}}
\end{figure*}
After all agents have made a decision and accumulated their corresponding payoff, they undergo a round of strategy revision phase, i.e., they can change their strategy from cooperation to defection, or vice versa.
In doing so, the population evolves until it reaches a final equilibrium (or steady-state).
The synergy factor plays a key role in this dynamics~\cite{perc03}, as it promotes cooperation. Notably, it is possible to compute, for a given topology (e.g. square lattice), the minimal value of $r$, here denoted as $r_{m}$, for which any $r>r_m$ allows cooperators to survive, or even to dominate.
From a statistical physics perspective, in particular referring to the Curie-Weiss model~\cite{wolski01}, we can now identify two different phases (or equilibria) ~\cite{huang01}: a paramagnetic equilibrium in which we observe the coexistence of cooperators and defectors, and a ferromagnetic equilibrium, implying that one species prevails.
Typically, the basic dynamics of the PGG let agents change their strategy according to rules based on the payoff~\cite{tomassini01}, e.g. agents imitate their richest neighbor, which is our definition of rational thinking~\cite{javarone07}.
In our model we aim to investigate the outcomes of the PGG in heterogeneous populations, i.e. composed of fitness-driven agents (FDAs) and conformity-driven agents (CDAs). Specifically, we map the fitness to the agents's payoff, so that the richest one is the fittest one. This assumption is far from general, as the fitness might be related to various factors, but we believe to be anyhow characterizing several realistic economic systems. 
Thus, in our model, FDAs tend imitate the richest neighbors, while CDAs tend imitate the most adopted strategy. So the imitation process, in particular for CDAs, depends on the local connectivity of the underlying interaction structure (i.e. the adopted topology).
As result our population is composed of $N = N_f + N_c$ agents, with $N_f$ is the number of FDAs and $N_c$ that of CDAs. Thus, we can introduce $\rho_f=N_f/N$ and $\rho_c=N_c/N$ to identify the density of FDAs and CDAs, respectively.
For the sake of clarity, we use the convention in which upper indices refer to the strategy (i.e. cooperation and defection), while lower indices to the agent's nature (i.e. conformity-driven and fitness-driven).
Both FDAs and CDAs change strategy by a stochastic rule. In particular, we implement a Fermi rule~\cite{perc03} to compute the transition probability between two different strategies for FDAs, so that the $y$-th agent imitates the $x$-th one, which reads
\begin{equation}\label{eq:fermi_function}
W(s^y \leftarrow s^x) = \left(1 + \exp\left[\frac{\pi^y - \pi^x}{K}\right]\right)^{-1}
\end{equation}
\noindent where $s^x$ and $s^y$ indicate the strategy of the agents $x$ and $y$, respectively, meanwhile $\pi^x$ and $\pi^y$ indicate their payoff; the constant $K$ parametrizes the uncertainty in adopting a strategy. By using $K = 0.5$, we implement a rational and meritocratic approach during the strategy revision phase~\cite{perc03}.
CDAs adopt a simple majority voting~\cite{galam01} rule to decide their next strategy: an agent computes the transition probability according to the density of neighbors having the strategy of majority. In doing so, FDAs act rationally, while CDAs follow a social behavior (i.e. conformism). 
Following the prescription of~\cite{perc03}, we arrange agents in a bi-dimensional regular lattice of degree $4$ with periodic boundary conditions (a torus).
Summarizing, our population evolves according to the following steps:
\begin{enumerate}
\item At $t = 0$, set an equal number of cooperators and defectors, and the density of conformists $\rho_c \in [0,1]$;
\item select randomly one agent $x$, and select randomly one of its neighbors $y$; 
\item each selected agent plays the PGG with all its five communities, then computes its payoff;  
\item agent $y$ performs the strategy revision phase according to its nature;
\item repeat from $(2)$ until an ordered phase is reached, or up to a limited number of time steps elapsed.
\end{enumerate}
We remark that the neighborhood for each agent has always $4$ agents. Therefore, one agent plays in $5$ different groups at a time, all composed of $5$ members.
Finally, we remind that agents may change strategy, i.e. from cooperation to defection (and vice versa), but they cannot change their nature (i.e. fitness-driven and conformity-driven). Although in real social systems individuals might change also their behavior (e.g. from CDA to FDA), in this work we aim to analyze the relation between the density of CDAs and the outcomes of the PGG. Therefore, we need to assume agents keep constant their behavior.
\begin{figure*}[t]
\centering
\includegraphics[width=0.75\textwidth]{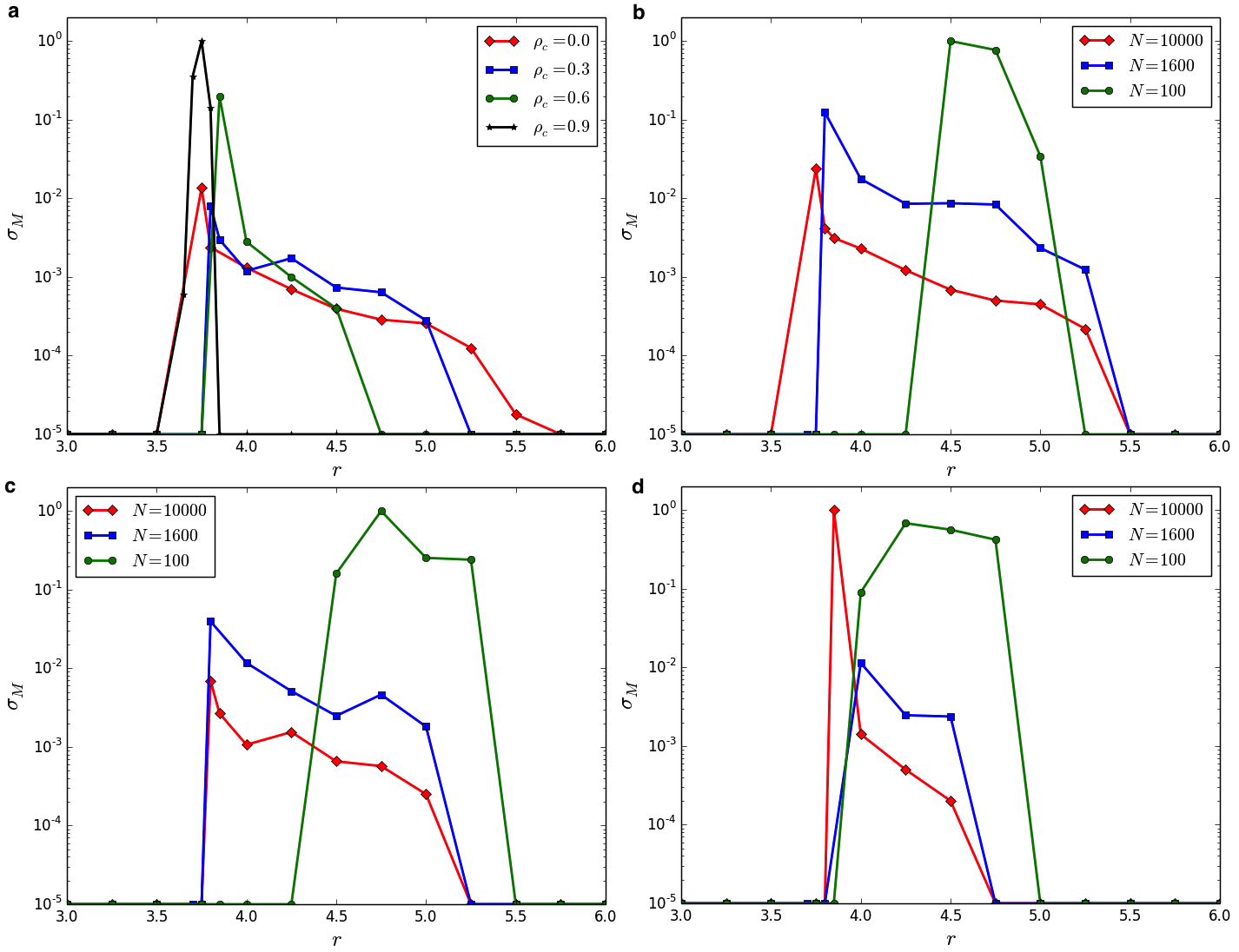}
\caption{\small (Color online) Variance ($\sigma_M$) of the order parameter $M$ as a function of the synergy factor $r$, for different configurations: \textbf{a} $N = 10^4$. \textbf{b} $\rho_c = 0.0$. \textbf{c} $\rho_c = 0.3$. \textbf{d} $\rho_c = 0.6$. Since we adopted a logarithmic scale for the $y$-axis, we highlight that all values equal to $10^{-5}$ correspond to $0$.\label{fig:phase}}
\end{figure*}

We investigate the behavior of the proposed model for different values of $\rho_c$, from $0$ to $1$, and of $r$. The latter assumes values in the range $[3,6]$ since, in this topology, it is known from~\cite{perc03} that the two thresholds for different equilibria are $r_m = 3.74$ and $r_M = 5.49$ in a FDA population. 
The threshold $r_m$ indicates that lower values of $r$ lead the population towards a phase of full defection at equilibrium. For intermediate values of $r$, i.e. $r_m \le r \le r_M$, the population reaches a disordered phase, i.e. a mixed phase characterized by the coexistence of both species at equilibrium; eventually, for values of $r>r_M$ cooperators succeed, i.e. the population reaches an ordered phase of full cooperation. 
In order to investigate the proposed model, we perform numerical simulations with populations of different size, from $N = 10^{2}$ to $N = 10^{4}$.
The first analysis is related to the distribution of strategies, at equilibrium, on varying the synergy factor $r$ and the density of conformists $\rho_c$ --- see figure~\ref{fig:cooperation}.
It is worth noting that the disordered phase becomes narrower as $\rho_c$ increases. Notably, we observe that $r_m$ and $r_M$ are strongly affected by $\rho_c$. At a first glance, as also reported in~\cite{perc04}, conformism fosters cooperation, as $r_M$ strongly reduces while increasing $\rho_c$.
On the other hand, for $\rho_c = 1$ a bistable behavior is expected as agents change strategy without considering the payoff.
In particular, the minimal threshold of synergy factor to avoid cooperators disappear reduces to values smaller than $r = 3.75$ (computed in~\cite{perc03} and confirmed here for $\rho_c = 0.0$) when $\rho_c$ is greater than $0.85$. 
Moreover, considering the higher threshold $r_M$ (i.e. that to obtain full cooperation for $\rho_c = 0.0$), we observe that even with low density of conformist agents, $r_M$ decreases, up to reach a value slightly smaller than $4.0$.
In the range $\rho_c \in [0.9, 1.0]$, a closer look allows to note a richer behavior of our model --- see plot \textbf{b} of figure~\ref{fig:cooperation}. We notice that defectors succeed only for values of $r$ smaller than $3.6$, while cooperators succeed for values of $r$ greater than $3.78$. As result the mixed phase is obtained only in a narrow range between the two listed values (i.e. $3.6 \le r \le 3.78$).
For values of $\rho_c \ge 0.97$ a bistable behavior can be observed: sometimes cooperators succeed, while other times fail (i.e. defectors succeed). Thus, since our results are computed as average values of different simulation runs, the colors represented in both plots of figure~\ref{fig:cooperation} in some cases reflect the probability to find the final population in a given status starting with those initial conditions (i.e. $r$ and $\rho_c$).

In order to characterize the transition at fixed $\rho_c$, since we observe qualitatively different phases, we tentatively try to identify the transition lines by studying the behavior of the variance as a function of $r$, which as we will see play the role of inverse `temperature'.
Here, the variance $\sigma_M$ is referred to the magnetization of the system~\cite{mobilia01}, which we identify as our order parameter, and given by
\begin{equation}\label{eq:magnetization}
M = \frac{1}{N}\sum_{i= 1}^{N} s_i
\end{equation}
\noindent with $s_i$ strategy of the $i$-th agent, i.e. $s =\pm 1$. Hence, the variance $\sigma_M$ is computed numerically, but can be easily identified as the susceptibility of the order parameter $\chi$,
\begin{equation}\label{eq:variance}
\sigma_M = \frac{1}{Z}\sum_{i= 1}^{Z} (M_i - \langle M\rangle)^2\equiv \chi
\end{equation}
\noindent with $Z$ number of simulations performed under the same conditions (i.e. fixed $r$ and $\rho_c$) and $\langle M\rangle$ average magnetization (computed in the same conditions).
Plot \textbf{a} of figure~\ref{fig:phase} (a) shows the variance $\sigma_M$ for different values of $\rho_c$: $0, 0.3, 0.6, 0.9$, as a function of the synergy factor $r$. 
As expected, we found that for $\rho_c=0.0$ the variance is maximum at $r_m \sim 3.75$. Plots \textbf{b, c, d} of figure~\ref{fig:phase} illustrate how these curves scale as we increase the number of agents for $\rho_c = 0.0$, $\rho_c=0.3$ and $\rho_c = 0.6$, respectively. We observe that in the case $\rho_c=0.6$, the limit $N\rightarrow \infty$ is critical, i.e. we find that there seem to exist a $r_{crit}$ for which $\lim_{N \rightarrow \infty} \chi_{N} \equiv \chi \approx (r-r_{crit})^{-\alpha}$ for some exponent $\alpha>0$. The universality class will be studied elsewhere.

\begin{figure*}[t]
\centering
\includegraphics[width=0.84\textwidth]{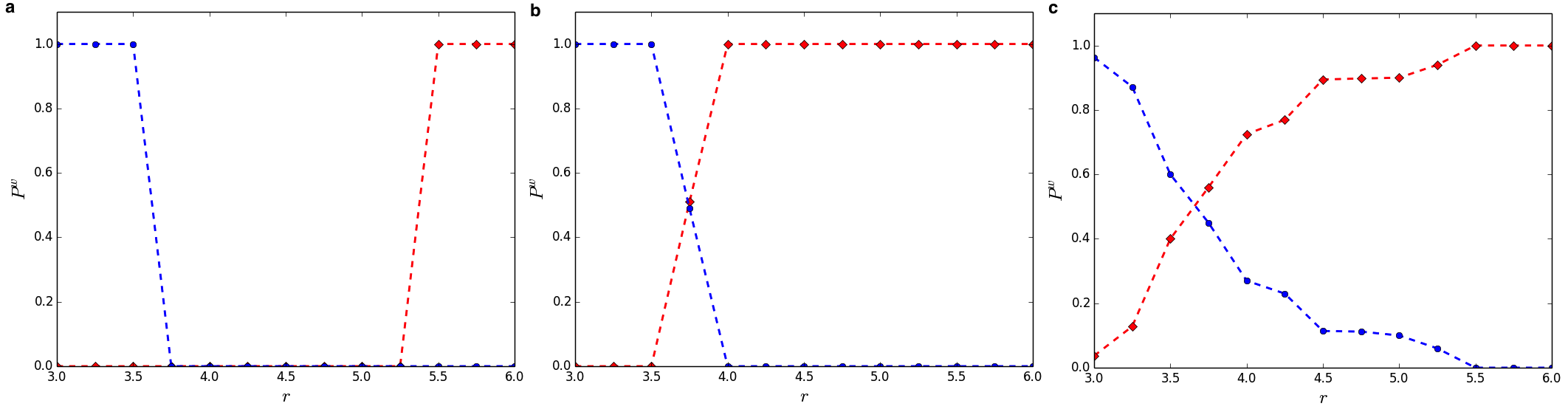}
\caption{\small (Color online) Probability to succeed as a function of the synergy factor $r$, in a population with $N = 10^4$, for the two species: cooperators, i.e., red dotted line (Diamonds $\Diamond$), and defectors, i.e., blue dotted line (Circles $\circ$). \textbf{a} $\rho_c = 0.0$. \textbf{b} $\rho_c = 0.9$. \textbf{c} $\rho_c = 0.99$. \label{fig:probability_r}}
\end{figure*}

\begin{figure*}[t]
\centering
\includegraphics[width=0.84\textwidth]{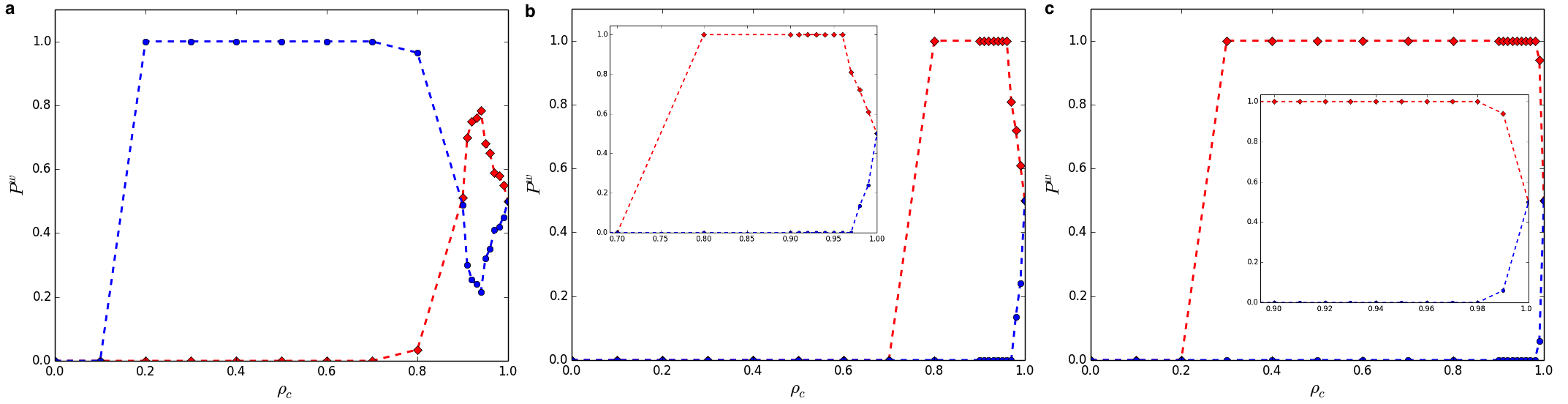}
\caption{\small (Color online) Probability to succeed as a function of the density of conformists $\rho_c$, in a population with $N = 10^4$, for the two species: cooperators, i.e., red dotted line (Diamonds $\Diamond$), and defectors, i.e., blue dotted line (Circles $\circ$). \textbf{a} $r = 3.75$. \textbf{b} $r = 4.0$. \textbf{c} $r = 5.25$. \label{fig:probability_rho}}
\end{figure*}

Then, in order to characterize the bistable behavior shown in Figure~\ref{fig:cooperation}, we study the probability for the system of being in the defecting or in the cooperating phase at the end of the simulation, as a function of $r$ (see figure~\ref{fig:probability_r}) and of $\rho_c$ (see figure~\ref{fig:probability_rho}).
In figure~\ref{fig:probability_r} the two dotted lines refer to the winning probabilities of defectors (i.e., blue) and of cooperators (i.e., red). Therefore, for $\rho_c = 0.0$, the two curves are zero in the intermediate range of $r$, i.e., $3.75 \le r \le r_M$, as none is expected to completely succeed.
Remarkably, increasing $\rho_c$ we found a decreasing paramagnetic range of $r$, disappearing for values $\rho_c \ge 0.8$. As shown in plots \textbf{b} and \textbf{c} of figure~\ref{fig:probability_r}, at least one curve is always greater than zero. 
Although we are dealing with success probabilities, it is worth noting that the summation of values taken by the two curves has to be $\le 1$, thus even zero as it means that none succeeds once the disordered phase is reached.
Moreover, figure~\ref{fig:probability_r} allows to observe the emergence of a bistable behavior, e.g. for $\rho_c = 0.9$ at $r = 3.75$ we have both curves having the same $P^w$, i.e., about $50\%$ of cases defectors prevail, while in the remaining cases cooperators succeed.
Figure~\ref{fig:probability_rho} aims to characterize the same bistable behavior on varying $\rho_c$ and keeping fixed $r$.
Plot \textbf{a} of Figure~\ref{fig:probability_rho} refers to $r = 3.75$ and it lets emerge an interesting result: in the range $0.2 \le \rho_c \le 0.7$ defectors prevail. This indicates that in this region conformism promotes defection, being $0$ the expected value of $P^w$ for both species. Moreover, the bistable behavior emerges as $\rho_c \ge 0.8$.
Plot \textbf{b} of figure~\ref{fig:probability_rho} refers to $r = 4.0$ and shows that the upper bound of the paramagnetic phase (i.e., $r_M$) is reduced to $4.0$ as $\rho_c \ge 0.8$. Then, a bistable behavior emerges for $\rho_c \ge 0.92$.
Eventually, in plot \textbf{c} of figure~\ref{fig:probability_rho} referred to  $r = 5.25$, we see that even for lower values of $\rho_c$ cooperators succeed, and the bistable behavior emerges for $\rho_c \ge 0.93$.
In the light of these results, we can state that when $r$ is close to the lower bound of the paramagnetic phase, i.e., $r_m$, conformism supports defection until the emergence of a bistable behavior. While, for higher values of $r$, conformism supports cooperation, and only for high values of $\rho_c$ the system becomes bistable.
\begin{figure}[h!]
\centering
\includegraphics[width=0.18\textwidth]{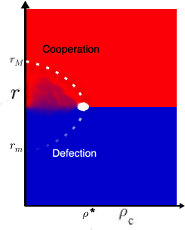}
\caption{(Color online) For $r_m<r<r_M$ and $\rho_c<\rho^*$, we observe a phase where defection and cooperation coexist, represented in the dashed white line, where the variance is continuous. For $\rho_c>\rho^*$, the transition from defection to cooperation is sharper at $r_c$. This picture gives the idea of the existence of a triple point at $r=r_m$, $\rho_c=\rho^*$ where three different behaviors coexist. }
\label{fig:critdiag}
\end{figure}

Finally, we construct an approximate phase diagram of our system --- see figure~\ref{fig:critdiag}. In the top part we have the domination of cooperation (i.e., red), and in the lower one that of defection (i.e., blue). Along the line separating the two parts above identified (at fixed $r$), we find an important point indicated as $\rho^*$ below described.
In an area of the left diagram between defection and cooperation, for $\rho_c<\rho^*$ and $r_m<r<r_M(\rho_c)$, defectors and cooperators coexist, with the prevalence of the former. In this region, it is possible to change the parameters to reach smoothly the cooperation region.
For $\rho_c>\rho^*$ we have the coexistence of cooperation and defection on the transition line $r=r_m$, due to the fact that $r_M$ approaches $r_m$ as an increasing function of $\rho_c$. The point in which $r_M=r_m$ is a triple point.

Summarizing, we studied the spatial PGG in the presence of social influences. In particular, we considered two different kinds of agents: conformity-driven agents (CDAs) and fitness-driven agents (FDAs). The former are those that tend imitate the strategy of majority, while the latter are those that tend to imitate the richest players.
In both cases, CDAs and FDAs update their strategy by considering only their neighborhood.
Previous studies~\cite{perc04,pena01,javarone06} reported that social influences strongly affect evolutionary games.  
The proposed model is different from those implemented in previous investigations (e.g.~\cite{perc04,javarone06}), results are similar and, on a quality level, further extends their findings.
Here, we highlight the prominent role of conformism in the spatial PGG: it seems that this social influence may lead the population towards different phases and behaviors, as full cooperation and bistable equilibria. In particular, conformism promotes the population to reach an ordered phase, even when a disordered one is expected.
For intermediate densities of conformists (e.g. $0.5$), the final equilibrium is that closer to that one would expect considering only FDAs, at a given $r$. Therefore, our investigations suggest that conformism drives the system towards ordered states, with a prevalence for cooperative equilibria.
Eventually, we focused on the identification of an order-disorder transition~\cite{huang01,moreno02,szabo01,galam03} that characterizes the behavior of our population on varying the degrees of freedom. 
To conclude, we found that the spatial PGG under social influences has a very rich behavior, characterized by different final states. 
\section*{Acknowledgements}
A.A. gratefully acknowledges financial support by the Swiss National
Science Foundation (P2LAP1-161864).
F.C. acknowledges support from Invenia Labs.

\end{document}